\documentclass[aps,preprint,longbibliography]{revtex4-1}
\usepackage{amssymb}
\usepackage{hyperref}
\usepackage{amsmath}
\usepackage{bm}
\usepackage{color}
\usepackage{epsfig}
\usepackage{float}
\usepackage{graphicx}

\begin{document}

\title{Time evolving matrix product operator (TEMPO) method
in a non-diagonal basis set based on derivative of the 
path integral expression}

\author{Shuocang Zhang}
\author{Qiang Shi}\email{qshi@iccas.ac.cn}
\affiliation{Beijing National Laboratory for
Molecular Sciences, State Key Laboratory for Structural Chemistry of
Unstable and Stable Species, Institute of Chemistry, Chinese
Academy of Sciences, Zhongguancun, Beijing 100190, China}
\affiliation{University of Chinese Academy of Sciences, Beijing 100049, China}

\begin{abstract}

The time-evolving matrix product operator (TEMPO) method is
a powerful tool for simulating open system quantum dynamics.
Typically, it is used in problems with diagonal system-bath
coupling, where analytical expressions for discretized
influence functional are available. In this work, we aim to
address issues related to off-diagonal coupling by extending
the TEMPO algorithm to accommodate arbitrary basis sets. The
proposed approach is based on computing the derivative of
the discretized path integral expression of a generalized 
influence functional when increasing one
time step, which yields an equation of motion valid for
non-diagonal basis set and arbitrary number of non-commuting
baths. The generalized influence functional is
then obtained by integrating the resulting differential
equation. Applicability of the the new method is then tested by
simulating one- and two- qubit systems coupled to both
$Z$- and $X$-type baths.

\end{abstract}

\maketitle

Quantum dynamics in open systems\cite{breuer02,weiss12}
represents a fascinating frontier in quantum physics, with
critical applications in diverse fields ranging from quantum
information to charge and energy transfer in molecules. A
major challenge in the theoretical treatment of open system
quantum dynamics lies in accurately capturing non-Markovian
effects and moving beyond commonly used second-order
perturbative treatments of the system-bath coupling, which
are important in many
problems\cite{hennessy07,rivas20,segal23}. In the
literature, advanced theoretical frameworks and methods have
been developed to address these challenges, including the
quasi-adiabatic path integral
(QUAPI)\cite{makarov94,makarov95,sim97}, hierarchical
equations of motion (HEOM)\cite{tanimura89,tanimura20}, and
methods based on tensor network
approaches\cite{bulla08,shi18,fux21,bose22}.

QUAPI is a powerful method for simulating non-Markovian
quantum dynamics\cite{makri95a,makri95b}. In recent years,
several new algorithms have been developed to enhance the
efficiency of QUAPI
calculations\cite{makri20,kundu20,strathearn18,richter22}.
Notably, the time-evolving matrix product operator (TEMPO)
method\cite{strathearn18} has significantly improved
efficiency in treating long memory effects by employing
matrix product state (MPS) techniques to reduce the
computational cost. Building on this, the process tensor
framework based on TEMPO (PT-TEMPO)\cite{jorgensen19} was
introduced, enabling the construction of a discretized
influence functional in the MPS form that can be reused for
time-dependent simulations, further enhancing computational
efficiency. Both TEMPO and PT-TEMPO methods have been
successfully applied to a variety of complex quantum
systems, including cavity polaritons\cite{fowler22} and spin
chains\cite{fux23}.

The TEMPO method\cite{strathearn18} utilizes the QUAPI
expression for the time-discretized path
integral\cite{makarov94,makri95a,makri95b}, which is
typically derived using a basis set consisting of the
eigenstates of the system operator that couples to the
collective bath coordinate\cite{makri95a}. Such an
expression is not readily available when working with a
non-eigenstate basis set of the system operator. This occurs
when the quantum system is coupled simultaneously to
multiple types of baths involving non-commuting system
operators.

This problem has been addressed in recent literature. For
example, Richter and Hughes introduced an additional set of
indices in the MPS representation to handle both diagonal
and off-diagonal operators\cite{richter22}. Additionally,
new methods have been developed that iteratively construct
and compress the influence functional using MPS
techniques\cite{ye21,cygorek22}, enabling simulations that
go beyond the commonly assumed linear coupling to a harmonic
bath and allowing for the simultaneous treatment of both
diagonal and off-diagonal system-bath couplings.

In this work, we propose an alternative method for handling
off-diagonal system-bath coupling based on the TEMPO
algorithm. The new approach involves deriving a differential
equation for the growth of a generalized influence
functional, which plays a role analogous to that of the
process tensor\cite{jorgensen19}. It is shown that 
this differential equation
enables the treatment of multiple bath couplings without the
need to introduce additional indices for non-commuting 
system operators. The generalized
influence functional in the MPS representation is then
obtained by integrating the differential equation. The
effectiveness of this method is demonstrated using 
one- and two qubit models coupled to $Z$- and $X$-type
baths, both individually and simultaneously.

We first consider a two level system that couples to a
dissipative environment (i.e., a spin-boson model), which is
described by the following Hamiltonian:
\begin{equation}
\label{eq-htot}
{H}_T={H}_{S}+{H}_{B}+{H}_{BS} \;\;.
\end{equation}
The system Hamiltonian $H_S$ in Eq. (\ref{eq-htot}) is give by:
\begin{equation}
\label{eq-hs}
{H}_{S} = \frac{\epsilon}{2} \sigma_z + \Delta \sigma_x \;\;,
\end{equation}
where $\epsilon$ and $\Delta$ are the energy 
bias and coupling constant between the $|0\rangle$ 
and $|1\rangle$ states.
The bath Hamiltonian $H_B$ and the system-bath 
interaction $H_{BS}$ term are given by:
\begin{equation}
\label{eq-hb}
{H}_{B} = \sum_{l=x,y,z}\sum_{j=1}^{N_B}
\left[\frac{p^2_{j,l}}{2m_j} +\frac{1}{2}m_j\omega^2_{j,l} 
q^2_{j,l} \right]\;\;,
\end{equation}
\begin{equation}
\label{eq-hsb}
{H}_{BS} = \sum_{l=x,y,z}\sum_{j=1}^{N_B}
c_{j,l}q_{j,l}\otimes {\sigma}_l\;\;.
\end{equation}
Here, ${\sigma}_{l}$ ($l=x,y,z$) represents the Pauli
operator. $p_{j,l}$, $m_j$, $\omega_{j,l}$,
$q_{j,l}$ denote the momentum, mass, frequency, and
coordinate of the $j$th 
harmonic oscillator mode
 of the bath. Eq.
(\ref{eq-hsb}) indicates that an independent linear
combination of the bath coordinates is coupled to the
$\sigma_x$,  $\sigma_y$, or $\sigma_z$ operators, resulting
in $X$-, $Y$-, or $Z$-type coupling to the bath,
respectively.

The system-bath interaction is characterized by the 
spectral density defined as:\cite{weiss12}
\begin{align}
J_l(\omega) = \frac{\pi}{2}\sum_{j}
\frac{c^2_{j,l}}
{\omega_{j,l}} \delta 
\left(\omega - \omega_{j,l} \right) \;\;.
\end{align}
We further assume that all $J_l(\omega)$s are the same
and can be described using the Ohmic spectral density 
with an exponential cutoff:
\begin{equation}
 J(\omega) =  2 \alpha \omega e^{-\frac{\omega}{\omega_{c}}} \;\;.
\end{equation}
In this work, we focus only on the $X$- and $Z$-type baths.

The initial state of the total system is assumed to be in a
factorized state: $\rho_{T} = \rho_S(0) \otimes e^{-\beta
H_B}$. In the QUAPI approach\cite{makri95a,makri95b}, the
total Hamiltonian is first partitioned into $H = H_S +
H_{{env}}$, where $H_{{env}} = H_B + H_{BS}$. The Trotter
decomposition is then utilized to divide the propagator into
discretized steps. We first consider the commonly studied
case of a single bath and employ diagonal basis functions,
where the basis set consists of eigenstates of the system
operator $X$ (either $\sigma_x$ or $\sigma_z$ in this work)
that couples to the collective bath coordinate. In this
case, the matrix element of the reduced density matrix at
time $t$ can be calculated as:
\begin{eqnarray}
\label{eq-diag-rho}
\langle x_2 | \rho_S(t) | x_1\rangle & = & 
\sum_{{\bf x}^{\pm} } \langle x_2 | 
e^{-\frac{i}{2\hbar}H_S\Delta t} | x_{N}^+ \rangle
\langle x_{N}^+ | e^{-\frac{i}{\hbar}H_S\Delta t} | x_{N-1}^+ \rangle
\cdots\langle x_1^+ | e^{-\frac{i}{2\hbar}H_S\Delta t} | x_0^+ \rangle
\langle x_0^+ | \rho_S(0) | x_0^- \rangle \nonumber \\
& & \langle x_0^- | e^{\frac{i}{2\hbar}H_S\Delta t} | x_1^- \rangle  \cdots
\langle x_{N-1}^- | e^{\frac{i}{\hbar}H_S\Delta t} | x_{N}^- \rangle
\langle x_{N}^- | e^{\frac{i}{2\hbar}H_S\Delta t} | x_1 \rangle
I({\bf x}^+,{\bf x}^-,\Delta t)\;\;,
\end{eqnarray}
where $\Delta t$ is the time step in the discrete path
integral expression, ${\bf x}^+ =
\left\{x_1^+,x_2^+\cdots,x_{N}^+\right\}$ and ${\bf x}^- =
\left\{x_1^-,x_2^-\cdots,x_{N}^-\right\}$ represent the
forward and backward paths. 

The influence functional is then obtained by integrating out all
the bath degrees of freedom (DOFs). In this process, the
operator $X$ in the ``quasi-adiabatic" environmental
Hamiltonian $H_{{env}}$ can be replaced by its
eigenvalues when computing the influence functional:
\begin{eqnarray}
\label{eq-diag-inf}
I({\bf x}^+,{\bf x}^-,\Delta t) & = & {\rm Tr}_{env} 
(e^{-\frac{i}{\hbar}H_{env}(x_{N}^+)\Delta t} \cdots 
e^{-\frac{i}{\hbar}H_{env}(x_1^+)\Delta t} \nonumber \\ 
& & \rho_B(0)e^{-\frac{i}{\hbar}H_{env}(x_1^-)\Delta t} 
\cdots e^{\frac{i}{\hbar}H_{env}(x_{N}^-)\Delta t}) \;\;.
\end{eqnarray}
The discretized influence functional for the harmonic bath 
can be calculated analytically\cite{weiss12}, which
is given by\cite{makarov94,makarov95,sim97}:
\begin{equation}
\label{eq-inf1}
I({\bf x}^+,{\bf x}^-,\Delta t) = 
e^{-\mathcal{F}({\bf x}^+,{\bf x}^-,\Delta t)} \;\;,
\end{equation}
\begin{equation}
\label{eq-inf2}
\mathcal{F}({\bf x}^+,{\bf x}^-,\Delta t) = \frac{1}{\hbar}
\sum_{k=0}^N\sum_{k'=0}^k
(x_k^+ - x_k^-)(\eta_{kk'}x_{k'}^+ - \eta_{kk'}^*x_{k'}^-) \;\;.
\end{equation}
Here, the coefficients $\eta_{kk'}$ are\cite{makarov94,makarov95,sim97}:
\begin{equation}
\label{eq-etakk'}
\eta_{kk'} = \int_{t_{k-1}}^{t_k}{\mathrm{d}t'}\int_{t_{k'-1}}^{t_k'}
{\mathrm{d}t''}C(t'-t'') \;\;,
\end{equation}
and
\begin{equation}
\label{eq-etakk}
\eta_{kk} = \int_{t_{k-1}}^{t_k}{\mathrm{d}t'}\int_{t_{k-1}}^{t'}
{\mathrm{d}t''}C(t'-t'') \;\;,
\end{equation}
where the bath correlation function is defined as:
\begin{equation}
\label{bathct}
C(t) = \frac{1}{\pi}\int_{0}^{\infty}{\mathrm{d}\omega}
{J(\omega)}{ \left[{\coth\left(\frac{\hbar\omega\beta}{2}\right)}{\cos\omega t}
-{i}{\sin\omega t} \right]} \;\;.
\end{equation}

Calculating the real-time path integral in Eq.
(\ref{eq-diag-rho}) becomes increasingly challenging for
long simulation times due to the summation over all possible
paths. For example, the real time Monte Carlo method suffers
from the sign problem, and can only be applied to short time
calculations\cite{egger94}. To address this problem, Makri and
coworkers developed a tensor-based method that takes
advantage of the short memory time of the bath correlation
functions in Eq. (\ref{bathct}), allowing the reduced
dynamics to be computed by propagating a tensor with a fixed
dimension\cite{makri95a,makri95b,sim97}. However, even with the
tensor based approach, computational costs grow rapidly as
the bath memory time or system size increases.

More recently, Strathearn {\it et al.}\cite{strathearn18}
proposed the TEMPO method, which utilizes MPS to represent
and compress the tensors involved in QUAPI calculations. As
an example, we consider using the TEMPO algorithm to
calculate the discretized influence functional. This
calculation differs slightly from the original TEMPO
approach\cite{strathearn18}, and has been employed to compute
the process tensor, as described in
Refs.\cite{fux21,fowler22,fux23}. For the influence
functional defined in Eq. (\ref{eq-inf1}), we denote its value
at the $N$th time step as $I_N(x_1^{\pm}, \cdots, x_N^\pm)$. The
influence functional at the $(N+1)$th time step can then 
be calculated as:
\begin{equation}
\label{IF recur}
I_{N+1}(x_1^{\pm},\cdots,x_{N+1}^\pm) = 
\Phi_{N+1}(x_1^{\pm},\cdots,x_{N+1}^\pm)  
I_{N}(x_1^{\pm},\cdots,x_{N}^\pm) \;\;,
\end{equation}
where the ``growth tensor" $\Phi_N$ is given by:
\begin{equation}
\label{eq-phiN}
\Phi_N({\bf x}^+,{\bf x}^-,\Delta t) = 
\exp\left( -\frac{1}{\hbar}\sum_{k=0}^N
(x_{N+1}^+ - x_{N+1}^-)(\eta_{N+1,k}x_{k}^+ 
- \eta_{N+1,k}^*x_{k}^-) \right)\;\;.
\end{equation}

It is noted that $\Phi_N$ contains the interaction between the
$(N+1)$th time step and all previous time steps, but not
interactions within the previous $N$ steps. In the TEMPO
method, both the influence functional $I_N$ and the ``growth
tensor" $\Phi_N$ are represented using MPS. For example, we can
write $I_N$ as:
\begin{equation}
I_{N} = \sum_{i_1,\cdots,i_N} B_{0}(i,i_1)
B_{1}(i_1,n_1,i_2)\cdots B_{N}(i_{N-1},n_N,i_N) 
B_{N+1}(i_N,j) \;\;.
\end{equation}

The key step now is the propagation from $I_N$ to $I_{N+1}$.
As shown in Ref.\cite{strathearn18}, $\Phi_N$ can be
conveniently written as a MPS with a bond dimension of $2
\times 2$. It is then multiplied with the MPS
representation of $I_N$ to obtain $I_{N+1}$. Subsequently,
the singular value decomposition (SVD)
method\cite{oseledets11} is applied to compress $I_{N+1}$
for use in the next step of the calculation. 

We now derive a differential equation approach to calculate
the influence functional $I_{N+1}$ from $I_N$. By starting
from the growth tensor in Eq. (\ref{eq-phiN}), we define a
new quantity depending on a parameter $\lambda$,
\begin{equation}
\label{eq-phiN-lambda}
\Phi_N^{\lambda}({\bf x}^\pm; \Delta t) = \exp\left( 
  -\frac{\lambda}{\hbar} \sum_{k=0}^N
(x_{N+1}^+ - x_{N+1}^-)(\eta_{N+1,k}x_{k}^+ - 
\eta_{N+1,k}^*x_{k}^-) \right) \;\;.
\end{equation}
Apparently, $\Phi_N^{(\lambda = 0)} = 1$, and  $\Phi_N^{(\lambda =1)} =\Phi_N$ 
in Eq. (\ref{eq-phiN}). We further define 
\begin{equation}
\label{eq-iflambda}
I^\lambda_{N+1}({\bf x}^\pm; \Delta t) = \Phi_N^\lambda 
I_N({\bf x}^\pm; \Delta t) \;\;.
\end{equation}
By taking the derivative of the above Eq. (\ref{eq-iflambda}) 
with respect to $\lambda$, we obtain:
\begin{eqnarray}
\label{eq-ifderiv}
\frac{\mathrm d}{\mathrm d\lambda}I^\lambda_{N+1}({\bf x}^\pm; 
\Delta t)
= \sum_k(x_{N+1}^+ - x_{N+1}^-)(\eta_{N+1,k}x_k^+ 
- \eta_{N+1,k}^*x_k^-) I^\lambda_{N+1}({\bf x}^\pm; 
\Delta t) \;\;,
\end{eqnarray}
with the initial condition $I_{N+1}^{(\lambda = 0)} =
I_{N}$. If we already know the influence functional $I_{N}$
at the $N$-th time step, we can then integrate Eq.
(\ref{eq-ifderiv}) with respect to $\lambda$ from 0 to 1 to
obtain the influence functional at the $(N+1)$-th time step,
$I_{N+1} = I_{N+1}^{(\lambda = 1)}$. 

We then show that the above approach can be extended to the
case of a non-diagonal basis set. For simplicity, we start
with the case of a single bath, as in the derivation of Eq.
(\ref{eq-ifderiv}), but using a general basis set that is not
necessarily the eigenstate of the system operator in
$H_{BS}$. To this end, the reduced density matrix is calculated as:
\begin{eqnarray}
\label{eq-offdiag1}
\langle s_2 | \rho_S(t) | s_1\rangle & = &
{\rm Tr}_{env} \Bigg( \sum_{{\bf s}^\pm_{1},{\bf s}^\pm_{2}}
\langle s_2 | e^{-\frac{i}{2\hbar}H_S\Delta t}| s_{N,1}^+ \rangle 
\langle s_{N-1,2}^+ | e^{-\frac{i}{\hbar}H_S\Delta t}| s_{N-1,1}^+ \rangle 
\cdots \langle s_{1,2}^+ | e^{\frac{i}{\hbar}H_{S}\Delta t} 
| s_{1,1}^+ \rangle \nonumber \\
& & \langle s^+_{0,2}|e^{-\frac{i}{2\hbar}H_S\Delta t}
|s^+_{0,1}\rangle
\langle s_{0,1}^+ | \rho_S(0)| s_{0,1}^- \rangle
\langle s_{0,1}^-| e^{\frac{i}{2\hbar}H_S\Delta t}| s_{0,2}^- \rangle 
\cdots \langle s_{N-1,1}^-| e^{\frac{i}{\hbar}H_{S}\Delta t}| 
s_{N-1,2}^- \rangle \nonumber \\
& & \langle s_{N,1}^-| e^{\frac{i}{2\hbar}H_{S}\Delta t}| 
s_{1} \rangle \tilde I_N({\bf s}^\pm_{1},{\bf s}^\pm_{2};\Delta t) \Bigg)  \;\;,
\end{eqnarray}
where ${\bf s}^\pm_{1} \ = \left \{|s^\pm_{0,1} \rangle, 
|s^\pm_{1.1} \rangle 
\cdots |s^\pm_{N,1} \rangle \right \} $ 
and ${\bf s}^\pm_{2} \ = \left \{|s^\pm_{0,2} \rangle, 
|s^\pm_{1,2} \rangle \cdots |s^\pm_{N-1,2} \rangle \right \} $ 
label the forward and backward paths.
We also require that  
$|s_1\rangle = |s^-_{N,2}\rangle$ and $|s_2\rangle =
|s^+_{N,2}\rangle$. 
Since $|s^\pm_{j,1}\rangle$ and $|s^\pm_{j,2}\rangle$ are no
longer eigenstates of the system operator $X$ in $H_{BS}$,
the number of indices doubles compared to Eq.
(\ref{eq-diag-rho}) for the diagonal basis set. Similar to
the case of diagonal basis set, the ``generalized influence
functional" $\tilde I_N({\bf s}^\pm_{1},{\bf s}^\pm_{2};\Delta t)$ 
is defined as:
\begin{eqnarray}
\label{eq-gif0}
\tilde I_N({\bf s}^\pm_{1},{\bf s}^\pm_{2};\Delta t) & = &
{\rm Tr}_{env} \Bigg( 
\langle s_{N,1}^+ | e^{-\frac{i}{\hbar}H_{env}\Delta t} 
| s_{N-1,2}^+ \rangle 
\langle s_{N-1,1}^+ | e^{-\frac{i}{\hbar}H_{env}\Delta t} | 
s_{N-2,2}^+ \rangle \cdots  
\nonumber \\
& & \langle s^+_{1,1}|e^{-\frac{i}{\hbar}H_{env}\Delta t}
| s^+_{0,2} \rangle
\rho_B(0)
\langle s_{0,2}^- | e^{\frac{i}{\hbar}H_{env}\Delta t} 
| s_{1,1}^- \rangle 
\cdots \nonumber \\
& & \langle s_{N-2,2}^- | e^{\frac{i}{\hbar}H_{env}\Delta t} 
|s_{N-1,1}^- \rangle 
\langle s_{N-1,2}^- | e^{\frac{i}{\hbar}H_{env}\Delta t} 
| s_{N,1}^- \rangle \Bigg)  \;\;.
\end{eqnarray}
It can be seen that, the above generalized influence function is
essentially equivalent to the process tensor used in
Refs.\cite{jorgensen19,cygorek22}.
To integrate out the harmonic bath DOFs, we insert again the 
diagonal basis set $|x^\pm_i\rangle$ at each time step $i$, 
\begin{eqnarray}
\label{eq-offdiag2}
\tilde I_N({\bf s}^\pm_{1},{\bf s}^\pm_{2};\Delta t)
& = &
{\rm Tr}_{env} \Bigg(\sum_{{\bf x}_N^\pm} 
\langle s_{N,1}^+ |  x_{N}^+ \rangle \langle x_{N}^+| 
e^{-\frac{i}{\hbar}H_{env}(x_{N}^+)\Delta t} | s_{N-1,2}^+ \rangle 
\langle s_{N-1,1}^+ |x_{N-1}^+ \rangle 
\nonumber \\ 
& &  \langle x_{N-1}^+|e^{-\frac{i}{\hbar}H_{env}(x_{N-1}^+)\Delta t} 
|s_{N-2,2}^+ \rangle \cdots \langle s^+_{1,1}| x_{1}^+ \rangle 
\langle x_{1}^+| e^{-\frac{i}{\hbar} H_{env}(x_{1}^+)\Delta t} 
| s^+_{0,2} \rangle 
\nonumber \\
& & \rho_B(0) 
\langle s_{0,2}^- | e^{\frac{i}{\hbar}H_{env}(x_{1}^-)\Delta t}| 
x_{1}^- \rangle \langle x_{1}^-| s_{1,1}^- \rangle \cdots  
\langle s_{N-2,2}^- | e^{\frac{i}{\hbar}H_{env}(x_{N-1}^-)\Delta t}| 
x_{N-1}^- \rangle \nonumber \\ 
& & \langle x_{N-1}^-|s_{N-1,1}^- \rangle 
\langle s_{N-1,2}^- | e^{\frac{i}{\hbar}H_{env}(x_{N}^-)\Delta t}|
x_{N}^- \rangle \langle x_{N}^-| s_{N,1}^- \rangle \Bigg)  \;\;,
\end{eqnarray}
and then integrate out the bath DOFs.
The generalized influence functional can then be calculated as:
\begin{equation}
\tilde I_N({\bf s}^\pm_{1},{\bf s}^\pm_{2};\Delta t)
 = \sum_{{\bf x}_N^\pm} 
\mathcal{I}_N({\bf s}^\pm_{1},{\bf s}^\pm_{2};{\bf x}_N^\pm; \Delta t) \;\;,
\end{equation}
where 
\begin{eqnarray}
\label{eq-gif}
\mathcal{I}_N({\bf s}^\pm_{1},{\bf s}^\pm_{2};{\bf x}_N^\pm; \Delta t) 
& = & \langle s_{N,1}^+ | x_N^+ \rangle
\langle x_N^+ | s_{N-1,2}^+ \rangle 
\langle s_{N-1,1}^+ | x_{N-1}^+ \rangle
\langle x_{N-1}^+ | s_{N-2,2}^+ \rangle  \nonumber \\
& & \cdots\langle s_{1,1}^+| x_1^+ \rangle 
\langle x_1^+ | s_{0,2}^+ \rangle
\langle s_{0,2}^- | x_1^- \rangle  
\langle x_1^- | s_{1,1}^- \rangle \nonumber \\
& & \cdots\langle s_{N-1,2}^-| x_{N}^- \rangle  
\langle x_{N}^- |s_{N,1}^-\rangle
e^{-\mathcal{F}({\bf x}^+,{\bf x}^-,\Delta t)} \;\;.
\end{eqnarray}
Here, ${\bf x}_N^\pm \ = \left\{x_1^\pm,x_2^\pm \cdots,x_{N}^\pm 
\right \}$.
As the expression for $e^{-\mathcal{F}({\bf x}^+, {\bf x}^-,
\Delta t)}$ is available in Eq. (\ref{eq-inf2}), the above
equation can be applied to perform TEMPO calculations in the
non-diagonal basis set, with the added complexity of
introducing a new set of indices ${\bf x}^\pm$. This
approach is very similar to the method used in
Ref.\cite{richter22}.

Since our goal is to perform calculations without relying on
the additional ${\bf x}^\pm$ variables, we apply the same technique
used to derive Eq. (\ref{eq-ifderiv}) to obtain a
differential equation for the generalized influence
functional in the non-diagonal basis set. To achieve
this, we first define the following quantity, which depends
on a parameter $\lambda$:
\begin{eqnarray}
\label{eq-giflambda}
\mathcal{I}^\lambda_{N+1}({\bf s}^\pm_{1},{\bf s}^\pm_{2};{\bf x}^\pm;\Delta t) 
& = & \langle s_{N+1,1}^+ | x_{N+1}^+ \rangle
\langle x_{N+1}^+ | s_{N,2}^+ \rangle
\langle s_{N,2}^-| x_{N+1}^- \rangle  
\langle x_{N+1}^- |s_{N+1,1}^-\rangle \nonumber \\
& & \Phi_N^\lambda 
\mathcal{I}_N({\bf s}^\pm_{1};{\bf s}^\pm_{2};{\bf x}^\pm;\Delta t) \;\;,
\end{eqnarray}
where $\Phi_N^\lambda $ is defined in Eq. (\ref{eq-phiN-lambda}).
 By taking the derivative over 
$\lambda$ in Eq. (\ref{eq-giflambda}), we obtain an equation that is similar to 
Eq. (\ref{eq-ifderiv}), but for the non-diagonal basis set:
\begin{eqnarray}
\label{eq-gifderiv1}
\frac{\mathrm d}{\mathrm d\lambda}\mathcal{I}^\lambda_{N+1}
({\bf s}^\pm_{1},{\bf s}^\pm_{2};
{\bf x}^\pm;\Delta t,\lambda) 
& = & \sum_k(x_{N+1}^+ - x_{N+1}^-)
(\eta_{N+1,k}x_k^+ - \eta_{N+1,k}^*x_k^-) \nonumber \\
& & \mathcal{I}^\lambda_{N+1}({\bf s}^\pm_{1},{\bf s}^\pm_{2};{\bf x}^\pm;\Delta t) \;\;.
\end{eqnarray}

The $\lambda$-dependent generalized influence functional can be 
defined as 
\begin{equation}
\tilde I^\lambda_N({\bf s}^\pm_{1},{\bf s}^\pm_{2};\Delta t)
 = \sum_{{\bf x}_N^\pm} 
\mathcal{I}^\lambda_N({\bf s}^\pm_{1},{\bf s}^\pm_{2};{\bf x}_N^\pm; \Delta t) \;\;,
\end{equation}
It can be seen that $\tilde{I}_{N+1}^{(\lambda =0)} = \tilde{I}_N $, and 
$\tilde{I}_{N+1}^{(\lambda =1)} = \tilde{I}_{N+1}$.

To obtain a closed equation of motion for $\tilde
I_{N+1}^\lambda ({\bf s}^\pm_{1}$, ${\bf s}^\pm_{2})$
without resorting to the ${\bf x}^\pm$ variables, we put Eq.
(\ref{eq-gif}) into the above equation. By further  
noticing that
\begin{eqnarray}
\langle s_{i,1} | x_i \rangle x_i = \langle s_{i,1} |X| x_i \rangle 
= \sum_{s'_{i,1}}\langle s_{i,1} | X | s'_{i,1}\rangle \langle s'_{i,1} 
| x_i \rangle \;\;,
\end{eqnarray} 
all the summation over $x_i$ can be incorporated 
into $I_{N+1}^\lambda({\bf s}^\pm_{1}$, ${\bf s}^\pm_{2}, \Delta t,\lambda)$, 
and we obtain:
\begin{eqnarray}
\label{eq-ifderiv2}
& & \frac{d}{d\lambda}\tilde I^\lambda_{N+1}({\bf s}^\pm_{1},{\bf s}^\pm_{2};
\Delta t)
\nonumber \\
& = & \sum_{s_{N+1,1}'^\pm}( \langle s_{N+1,1}^+ | X | s_{N+1,1}'^+ \rangle 
- \langle s_{N+1,1}'^- | X | s_{N+1,1}^- \rangle) 
\sum_k \sum_{s_{k,1}'^\pm}(\eta_{N+1,k} \langle  s_{k,1}^+ | 
X | s_{k,1}'^+\rangle  \nonumber \\ & & 
- \eta_{N+1,k}^* \langle s_{k,1}'^- | X | s_{k,1}^- \rangle)
\tilde I^\lambda_{N+1}(s_{0,1}^\pm,s_{1,1}^\pm,\cdots s_{k,1}'^\pm, 
\cdots;s_{N+1,1}'^\pm,
{\bf s}^\pm_{2};\Delta t)\;\;,
\end{eqnarray}
which is a closed form and does not contain the ${\bf
x}_N^\pm$ variables. It can be shown that, when $ |{\bf
s}^\pm_{1} \rangle$, $|{\bf s}^\pm_{2}\rangle$ are chosen as
eigenstates of the $X$ operator, $s_{k,1}^\pm =
s_{k,2}^\pm$, and the above equation reduces to the case of
the diagonal basis set in Eq. (\ref{eq-ifderiv}). 

The above approach can be extended to multiple bath problems. 
In this case, the contribution from 
each bath are just added up to give the following equation:
\begin{eqnarray}
\label{eq-ifderiv3}
& & \frac{d}{d\lambda}\tilde I^\lambda_{N+1}({\bf s}^{\pm}_{1},
{\bf s}^{\pm}_{2};\Delta t)
\nonumber \\
& = & \sum_{l=x,z} \sum_{s_{N+1,1}'^{\pm}}( 
  \langle s_{N+1,1}^{+} | \sigma_l  | s_{N+1,1}'^{+} \rangle 
- \langle s_{N+1,1}'^{-} | \sigma_l  | s_{N+1,1}^{-} \rangle) 
\sum_k \sum_{s_{k,1}'^{\pm}}(\eta^l_{N+1,k} \langle 
   s_{k,1}^{+} | \sigma_l  | s_{k,1}'^{+}\rangle 
\nonumber \\ & & 
- \eta_{N+1,k}^{l*} \langle s_{k,1}'^{-} | \sigma_l  
| s_{k,1}^{-} \rangle)
\tilde I^\lambda_{N+1}(s_{0,1}^{\pm},s_{1,1}^{\pm},\cdots 
s_{k,1}'^{\pm}, \cdots;s_{N+1,1}'^{\pm},
{\bf s}^{\pm}_{2};\Delta t)\;\;.
\end{eqnarray}

\begin{figure}
\centering
\includegraphics[width=14cm]{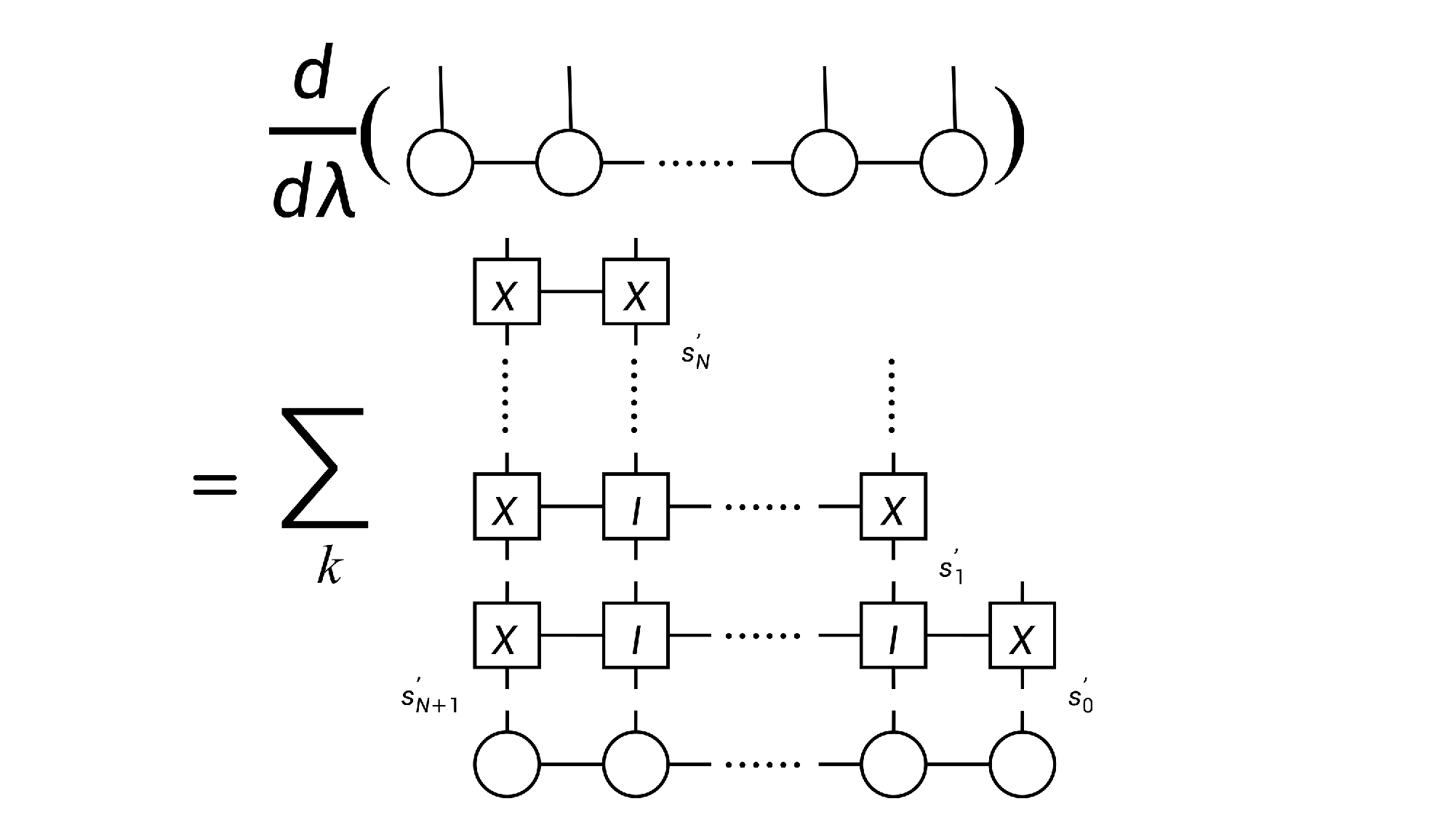}
\caption{Schematic view of the tensor network structure of
the differential equation for the general influence functional
in Eq. (\ref{eq-ifderiv3}). The circles indicate nodes of the 
generalized influence functional. In the square boxes, $X$ is 
the matrix representation of the corresponding system 
operator, and $I$ is the identity matrix.}
\vspace{30em}
\label{fig-schematic}
\end{figure}

Eq. (\ref{eq-ifderiv3}) is the main result of this paper.
Fig. \ref{fig-schematic} shows a schematic view of 
its structure in the ensor network representation.
The generalized 
influence functional in Eq. (\ref{eq-gif}) can then be 
computed by integration with respect to $\lambda$, 
which can be further utilized to obtain the reduced 
dynamics by using Eq. (\ref{eq-offdiag1}).

\begin{figure}
\centering
\includegraphics[width=14cm]{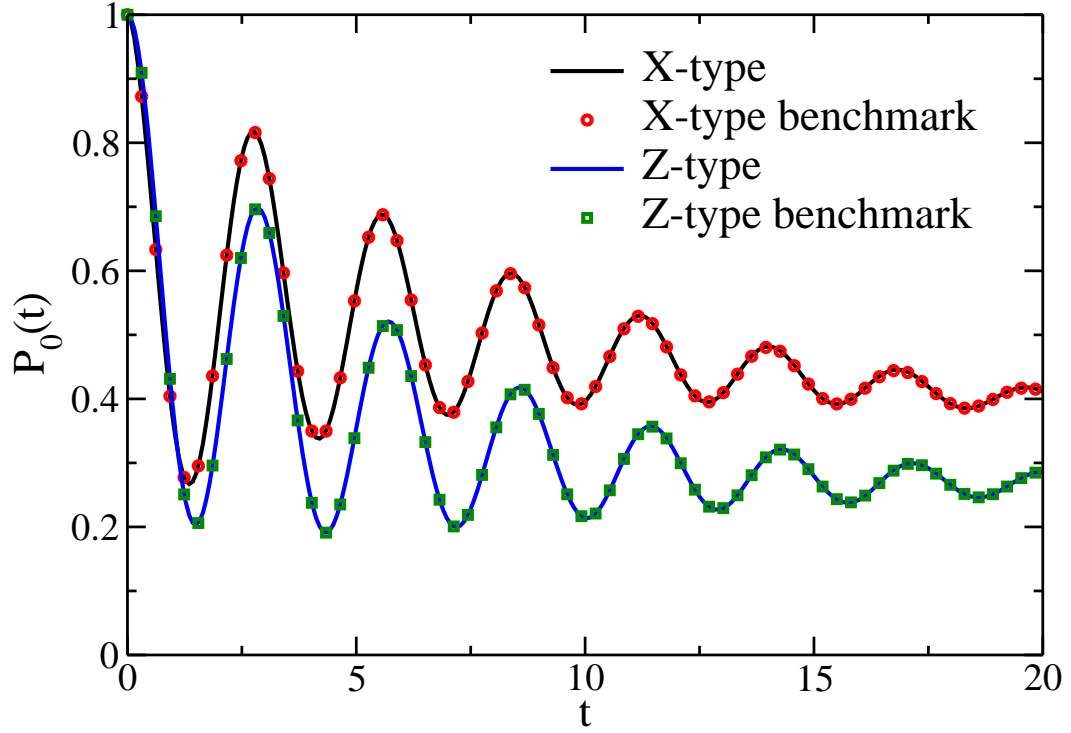}
\caption{ 
Population dynamics of a two level system
described by the Hamiltonian in Eqs. (1-6). The system is
coupled to either $X$-type (black) or $Z$-type (red) baths.
The solid curves represent results obtained using the
differential equation in Eq. (\ref{eq-ifderiv3}) with
non-diagonal basis sets, while the symbols correspond to
benchmark results from the conventional TEMPO method with diagonal
basis sets. See the main text for further details.
}
\label{1type}
\end{figure}

\begin{figure}
  \centering
  \includegraphics[width=14cm]{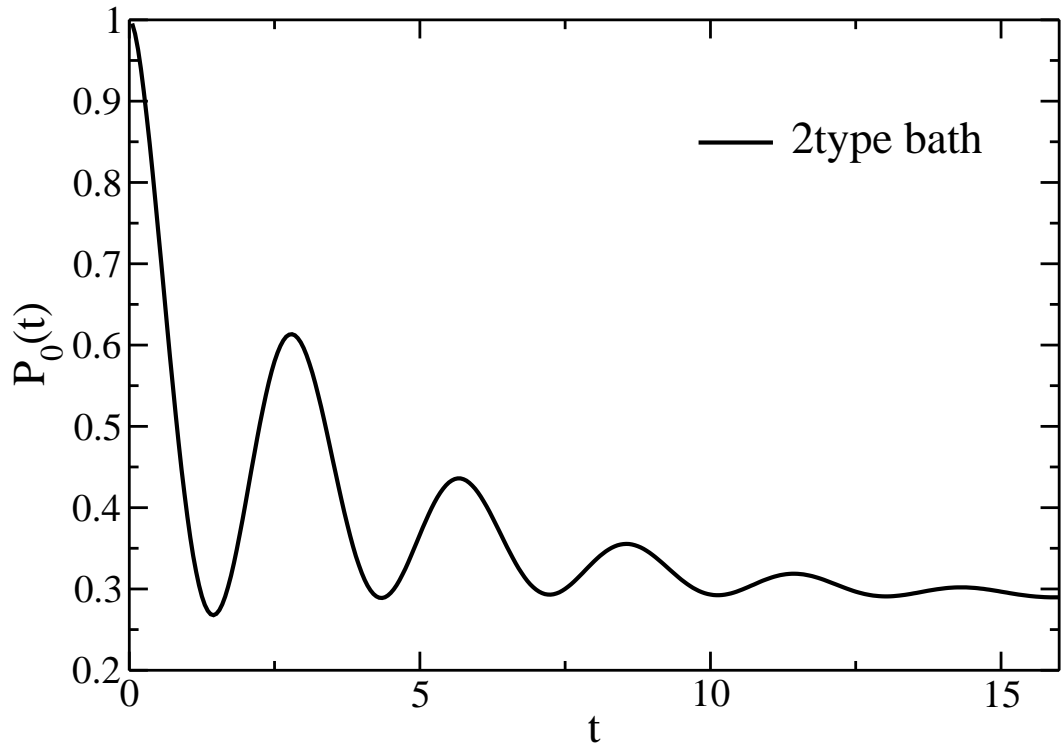}
  \caption{
Population dynamics of a two level system
coupled simultaneously to $X$- and $Z$-type baths.}
\label{fig-2type}
\end{figure}

We first use a one-qubit system with a single type of
system-bath interaction ($X$ or $Z$) to demonstrate that Eq.
(\ref{eq-ifderiv2}) produces the correct result using a
non-diagonal basis set. Fig. \ref{1type} shows the
population dynamics of the two level system coupled to $X$-
and $Z$-type baths. For the $X$-type bath, the eigenstates
of the $\sigma_z$ operator are used as the basis set, while
for the $Z$-type bath, the eigenstates of the $\sigma_x$
operator are used. The system is initially prepared in
the $|0\rangle$ state, and the parameters used in the
simulation are $\epsilon = 1.0$, $\Delta = 1.0$, $\alpha =
0.1$, $\omega_c = 5.0$, and $\beta = 2.5$. The standard 
TEMPO approach with diagonal basis set is used to 
calculate the benchmark results. In a second
example in Fig. \ref{fig-2type}, the population dynamics 
for a two level system coupled simultaneously to
both $X$- and $Z$-type baths is shown, where 
Eq. (\ref{eq-ifderiv3}) is used to obtain the generalized
influence functional.

\begin{figure}
\centering
\includegraphics[width=14cm]{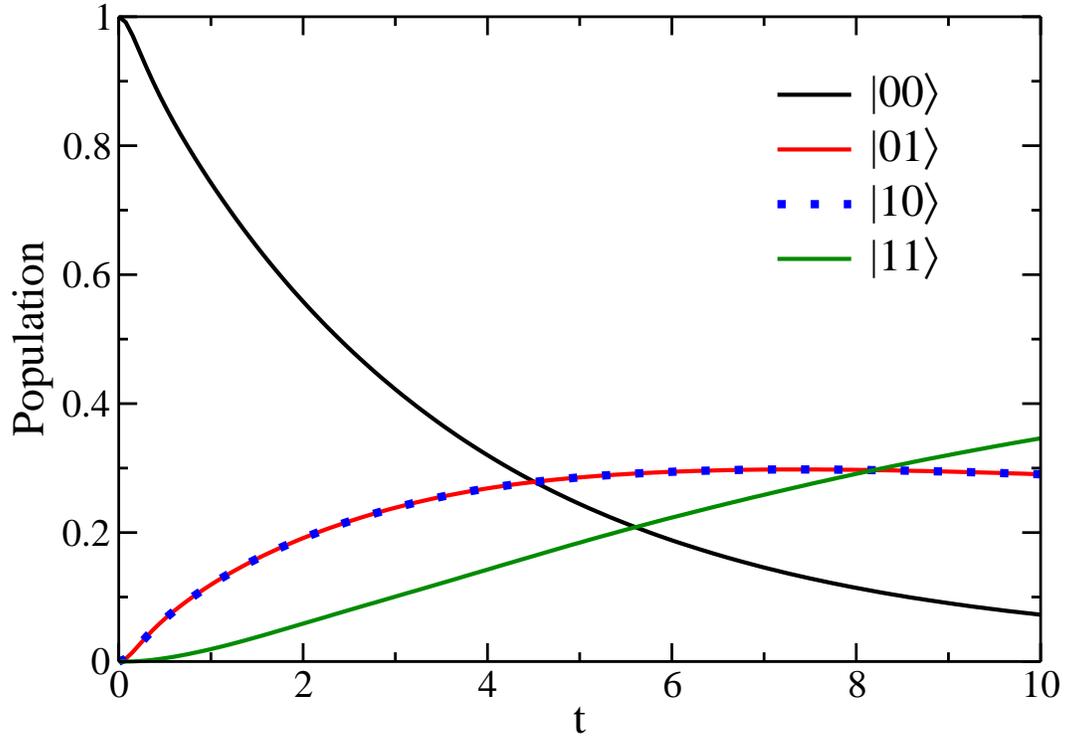}
\caption{
Population dynamics of a two-qubit system coupled
simultaneously to $X$- and $Z$-type baths. The initial state
is prepared in $|00\rangle$.
}
\label{fig-2bit}
\end{figure}

Finally, we consider a model of two qubits without internal
coupling ($\Delta = 0$), each coupled independently to its
own $X$- and $Z$-type baths, while at the same time, 
the one-qubit excited states are coupled via 
\begin{equation} H_{int} = J \left(
|01\rangle\langle 10| + |10\rangle\langle 01| \right) \;\;.
\end{equation} 
In this case, the generalized influence
functional is calculated in the same way as in the one-qubit
cases presented above, and is then used to perform the
two-qubit simulations in a way similar to the PT-TEMPO
method\cite{jorgensen19,cygorek22}. Results for the
population dynamics of the four states in the two-qubit
system are shown in Fig. \ref{fig-2bit}, with the same
parameters as those in Fig. \ref{fig-2type}, and $J = 1.0$.

In summary, we derive a differential equation to calculate
the generalized influence functional, as shown in Eq.
(\ref{eq-ifderiv3}). This new approach does not depend on
the specific choice of basis set and provides an efficient
solution to handle off-diagonal system-bath coupling and
non-commuting system-bath interactions within the TEMPO
framework. The proposed method is tested through simulations
of one- and two-qubit systems interacting with different
combinations of  $X$- and $Z$-type baths. It is expected
that the new approach could be useful in cases where the
quantum system is coupled simultaneously to multiple baths, 
or in cases where using
a non-diagonal basis might be advantageous.


\acknowledgments
This work is supported by NSFC (Grant Nos. 21933011 and 22433006).

\pagebreak

\pagebreak

\pagebreak

\pagebreak

\end{document}